\documentclass[12 pt]{article}
\usepackage{graphicx}
\usepackage{color}
\begin{document}
\title{\bf Simulating the Hyper-Kamiokande Detectors}
\author{\bf Srikanta Sinha\\
            3A, Sharda Royale Apt.,\\
            G. M.Palya, Bengaluru-560 075, INDIA.\\
	    email: sinha.srikanta@gmail.com}
\maketitle

\newpage

\section{\bf ABSTRACT}
   The Hyper-Kamiokande Water-Cherenkov Neutrino detectors are simulated
using Monte-Carlo and analytical methods. A few simple events are also
simulated and these preliminary results are presented.

\section{\bf INTRODUCTION}
   Neutrinos are the most elusive fundamental particles in the universe.
Contrary to popular notions their flux (at the surface of the earth) 
is very high. At the same time 
they interact with matter primarily through the weak interaction process
making their detection and study rather difficult. This is so since the
weak interaction cross-sections are much smaller (typ. $~ 10.0 E-37 cm^{-2}$)
compared to the strong interaction cross-sections (typ. $~ 10.0 E-25 cm^{2}$)
and electro-magnetic interaction cross-sections (typ. $~ 10.0 E-03 cm^{-2}$).
  Neutrinos are produced in very large numbers at the core of the Sun (during
the fusion of hydrogen into helium) and also during Supernova explosions 
that mark the death of massive stars. High energy cosmic rays while traveling towards
the earth's surface interact with the terrestrial atmospheric nuclei and 
produce neutrinos in these interactions.
  Large number of neutrinos are produced in nuclear reactors (used primarily
for energy production) and also in high energy particle accelerators.
  The mean energies of neutrinos and their spectral energy distributions produced in 
different natural and man-made processes are, however, quite different.
  Study of neutrinos help us in understanding the physics of the processes 
taking place in the core of the Sun, of neutrino oscilltions (the conversion
of a particular type of neutrino to another type, say, that of an electron-type
neutrino to a muon-type neutrino or say, a muon-type neutrino to a tauon-type
neutrino.  
    Due to the extremely small interaction cross-sections ($10.0 E{-37}cm^{-2}$) of neutrinos
extremely massive (hundreds of kilotons or megatons) detectors are required to
detect significantly large number of events. Also, another very important necessity
is that these detectors have to be located deep underground to significantly reduce
the very large number of high energy cosmic ray muons that incessantly bombard the
earth's surface.
\section{\bf THE DETECTOR SYSTEM}
   The details of the Hyper-Kamiokande underground detectors are available elsewhere [1].
Here we give only the basic configuration and information that are necessary for
simulating events in these detectors.
   The Hyper-K detector is located underground at a depth of 650 meters in the famous
Kamioka mines in Japan. The detector is basically a 'water cherenkov charged particle
detector'. Ultrarelativistic charged particles emit electromagnetic shock waves in the
ambient medium when their velocities exceed that of the speed of light in the medium.
These shock waves are emitted in the form of a cone with the moving charged particle
being at the vertex of the cone. This radiation is known as Cherenkov radsiation and
its wavelength is concentrated in the visible blue or near UV region of the electormagnetic
spectrum and may be detected easily using photosensors like photo-multiplier tubes or 
PMTs.
    The Hyper-K detector consists of a huge cylindrical chamber filled with ultrapure
water. The outer dimensions of the chamber is as follows: its height is 60 meters and its
diameter is 74 meters. 260,000 metric tons of water is required to fill the chamber.
    The wall of the Inner Detector (ID) is a cylinder having a height equal to 54.8
meters and a diameter of 70.8 meters. This Inner Detector (ID) is separated from the
optically isolated Outer Detector (OD) by a $60 cm$ wide dead space. The PMT high
voltage and electronics cables run through this dead space. The OD has a thickness of
1 meter in the barrel region and 2 m at the top and bottom of the Inner Detector.
A total of 39,424 PMTs cover the inside wall of the Inner Detector. These PMTs are
hemispherical,each having a diameter equal to $50 cm$ (Hamamatsu Type R12860-HQE).
They have a large quantum efficiency ($30\%$ at $ \lambda $ = 390 nm). 
The collection efficiency is $95 \%$ when the gain equals $10.^{7}$. The transit time 
spread for these PMTs is 2.7 nsec (FWHM).
   The total volume of the water is separated into two optically isolated detectors by
the mechanical structure that holds the PMTs. Thus there is an Inner Detector (ID) and
an Outer Detector (OD).
\section{\bf SIMULATING THE DETECTORS and EVENTS}
    Some details about the simulation methods and procedures are available
elsewhere [2], [3].
    The detector geometrical configuration (the inner and outer cylinders and
the top and bottom circular walls) including the positions of each PMT are
calculated using standard analytical formulae. 
    The total number of Cherenkov photons emitted in the wavelength band
$\delta$ $\lambda$ is calculated as follows:
\begin{equation}
 dN = 2 \pi \alpha dl (1 - 1/ n^{2} \beta^{2}) 
\\(1/\lambda_{1} - 1/\lambda_{2})
\end{equation}
    Here $\alpha$ is the fine structure constant, dl is the track length of the particle
and $\beta$ is the velocity of the particle in units of the velocity of light.
$\lambda_{1}$ and $\lambda_{2}$ define the boundaries of the wavelength band.
    All these photons are emitted along the surface of a cone having a vertex angle
\begin{equation}
 cos\theta = 1/(n\beta)
\end{equation}
    The point of emission of each photon along the particle's track is calculated
using a uniform random number. Using three-dimensional geometry, the path of each photon
is traced through the ambient medium and finally the hit-point on the detector's wall
is determined. Also, the time delay due to the propagation of each photon is calculated
from the point of its emission to the PMT.
    
   Another important aspect of the new MC code is that it is written almost 
entirely using the FORTRAN 90/95
language.
\section{\bf RESULTS}
     A computer simulated image of the HYPER-K cylindrical detector is shown in Fig.1.
Since the number of PMTs is too large, all the individual PMTs are not resolved
in the figure. The RED colored grid is the OUTER cylinder. The center
of each red colored PLUS sign  
is the center of an individual PMT.
The GREEN portion is the INNER cylinder. The DEEP BLUE semi-circle
is half of the TOP wall. Similarly, the PURPLE semi-circle is HALF of
the BOTTOM wall.
\begin{figure}
\centering
\includegraphics[width=9cm, height=8cm]{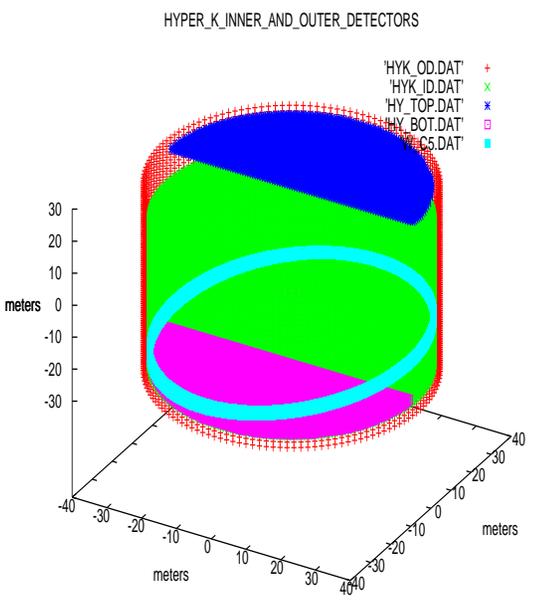}
\caption{The Hyper Kamiokande Inner and Outer Cylinders}
\end{figure}
    A simulated event is shown in Fig.2.
\begin{figure}
\centering
\includegraphics[width=9cm, height=8cm]{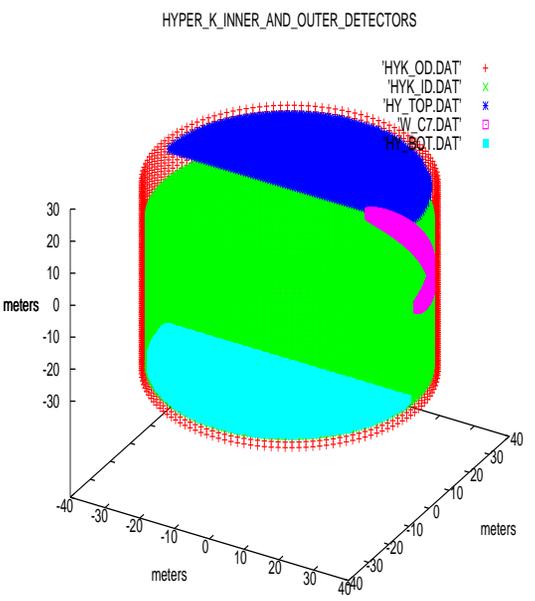}
\caption{A Typical Muon Event (simulated)}
\end{figure}
    Another simulated event is shown in Fig.3.
\begin{figure}
\centering
\includegraphics[width=9cm, height=8cm]{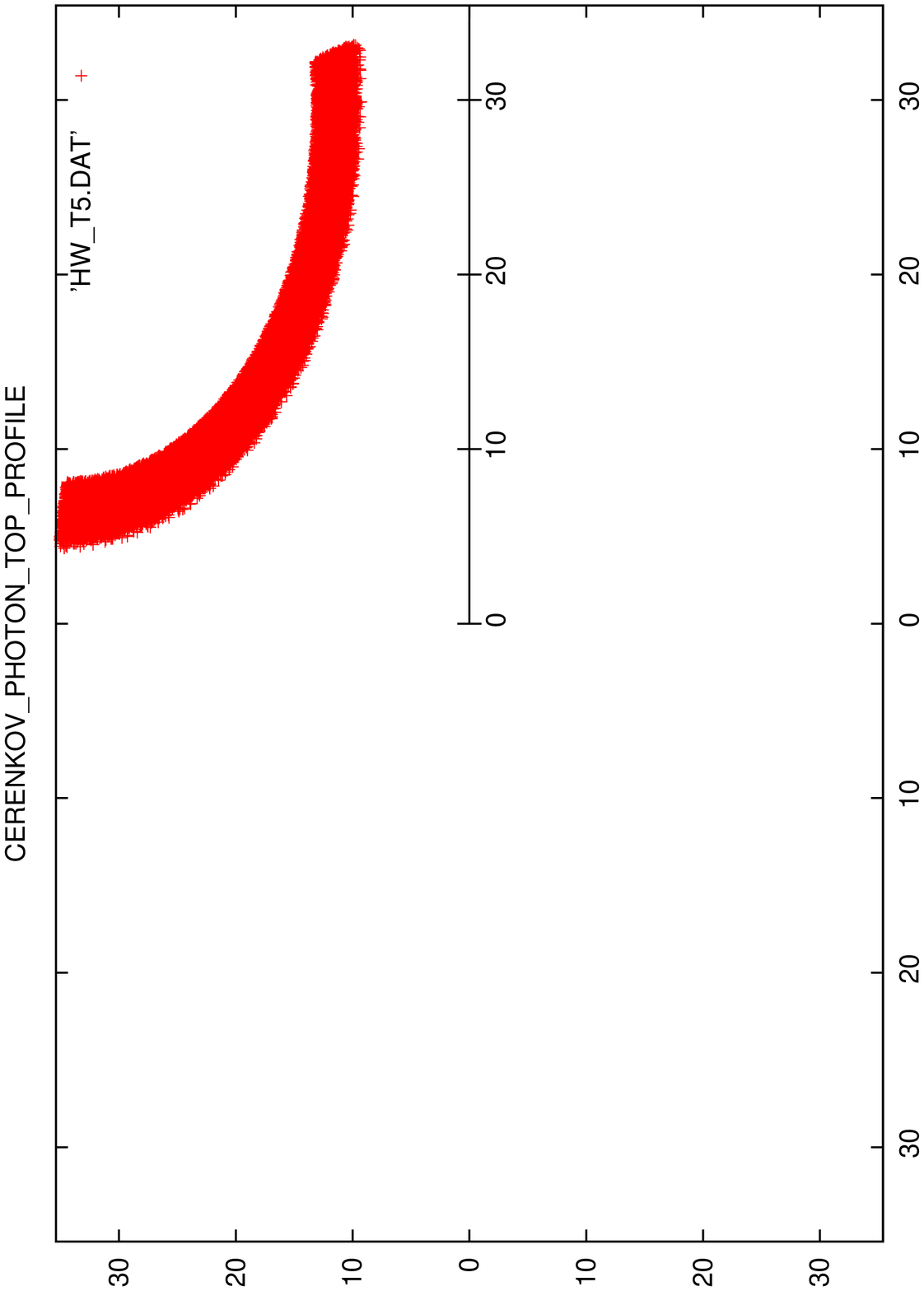}
\caption{A Second Event (HIT PMTs on the TOP Wall are displayed.)}
\end{figure}
\section{\bf DISCUSSIONS and CONCLUSION}
    Detailed physics have to be incorporated to make the
simulations much more realistic. Some of these processes have
already been included in the latest procedures and test runs taken.
Careful checking is required. These will help in getting
accurate estimates of physical parameters. The simulation
procedures are to be used to create different event topologies. 

\section{\bf ACKNOWLEDGEMENTS}
    I would like to express my deep sense of gratitude to all my
teachers, especially to Prof. A.K. Biswas and Dr. B.K. Chatterjee
who had been great sources of inspiration and support to me. I also
thank the HYPER-K collaboration to make lot of information available
about this great experiment on their website.


\begin{thebibliography}{3}
\bibitem{1}{KEK-PREPRINT-2016-21}
\bibitem{2}{Simulation of X-ray and gamma ray detector response and spectral
\\deconvolution-Sinha Srikanta, Advances in Space Research, Vol. 38,
\\Issue 12, 2006. p 3005-3007.}
\bibitem{3}{A High Energy Electron and Photon Detector 
\\Simulation System- Sinha Srikanta,
\\arXiv preprint arXiv: 0810.0459.}
\end{thebibliography}
\end{document}